\documentclass[multphys,vecphys]{svmult}

\newcommand{\be}[1]{\begin{equation} \label{(#1)}}
\newcommand{\ee}{\end{equation}}
\newcommand{\ba}[1]{\begin{eqnarray} \label{(#1)}}
\newcommand{\ea}{\end{eqnarray}}

\newcommand{\beq}{\begin{equation}}
\newcommand{\eeq}{\end{equation}}
\newcommand{\beqa}{\begin{eqnarray}}
\newcommand{\eeqa}{\end{eqnarray}}

\newcommand{\kf}{k_{\mathrm F} }
\newcommand{\Sigs}{\Sigma_{\mathrm s} }
\newcommand{\Sigv}{\Sigma_{\mathrm v} }
\newcommand{\Sigo}{\Sigma_{\mathrm o} }
\newcommand{\mst}{ {\tilde m}^* }
\newcommand{\mstq}{ {\tilde m}^{*2} }
\newcommand{\est}{ {\tilde E}^* }
\newcommand{\bfgamma}{\mbox{\boldmath$\gamma$\unboldmath}}
\newcommand{\pabs}{|{\bf p}|}

\usepackage{makeidx}     
\usepackage{graphicx}    
\usepackage{multicol}    

\makeindex             

\begin{document}

\title*{The Relativistic Dirac-Brueckner Approach to Nuclear Matter}
\titlerunning{Dirac-Brueckner Approach} 
\author{Christian Fuchs\inst{1}}
\institute{Institut f\"ur Theoretische Physik, 
Universit\"at T\"ubingen, D-72076 T\"ubingen, Germany}
%
%
\maketitle
An overview on the relativistic 
Dirac-Brueckner approach to the nuclear many-body
problem is given. Different approximation schemes are discussed, 
with particular emphasis on the nuclear self-energy and  
the saturation mechanism of nuclear matter. 
I will further discuss extensions of the standard approach, amongst other 
things the inclusion of non-nucleonic degrees of freedom,  
many-body forces and finally compare relativistic and non-relativistic 
approaches. 

\section{Introduction}
\begin{figure}
\begin{center}
\leavevmode
\includegraphics[width=.5\textwidth]{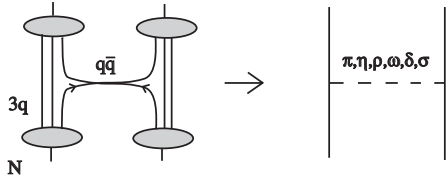}
\end{center}
\caption{Schematic representation of the One-Boson-Exchange 
model for the nucleon-nucleon interaction.
The figure is taken from \protect\cite{muether00}.}
\label{obep}
\end{figure}

An {\it ab initio} description of dense nuclear matter which is 
based on QCD as the fundamental theory of strong interactions is 
presently not possible and will not be in the foreseeable future. The 
reason lies in  the highly non-perturbative character of 
the formation of hadronic 
bound states and their interactions. Hence a quantitative 
description of nuclear many-particle systems  has 
to be based on effective theories. Particularly successful are 
theories which's effective degrees of freedom are 
hadrons, i.e. nucleons (and their excited states) and mesons. 
The nucleon-nucleon interaction 
is thereby described by the exchange of mesons 
as depicted in Fig. \ref{obep}. Modern One-Boson-Exchange Potentials 
(OBEP), as e.g. the Bonn potentials \cite{bonn2,bonn},  are usually 
based on the exchange 
of the six non-strange mesons: $\sigma$ (scalar, iso-scalar), 
$\omega$ (vector, iso-scalar), $\rho$ (vector, iso-vector),
$\pi$ (pseudo-scalar, iso-vector), $\eta$ (pseudoscalar, iso-scalar), 
$\delta$  (scalar, iso-vector). 

The connection of the hadronic to the QCD world is reflected in the 
{\it quark-hadron-duality}. Due to the almost vanishing masses of the 
light (current) quarks chiral symmetry of the QCD Lagrangian is 
almost fulfilled. It is, however, spontaneously broken by the 
large non-vanishing vacuum expectation values of the quark 
and gluon condensates which are responsible for the finite hadron 
masses. These condensates which are the basic quantities in the 
non-perturbative regime of QCD change dramatically in the medium. 
The connection to the energy density $\epsilon$ in terms of hadronic 
degrees of freedom is in principle given via the Hellmann-Feynman 
theorem 
\begin{equation}
\langle {\bar q}q \rangle_{\rho} =\langle {\bar q}q \rangle_{\rm vac} 
+ \frac{1}{2} \sum_{h} \frac{\partial \epsilon}{\partial h}
\frac{\partial m_h}{\partial m_q}   
\label{hell}
\end{equation}
where the sum runs over all hadronic contributions $h$ to the energy 
density $\epsilon$. However, neither this part can uniquely be fixed nor 
the second part of eq. (\ref{hell}), i.e. the derivatives of the 
hadron masses with respect to the current quark masses. Thus QCD can 
help to constrain hadronic theories but an exact mapping of the two 
worlds is still a dream for the future. 

\begin{figure}
\begin{center}
\leavevmode
\includegraphics[width=.4\textwidth]{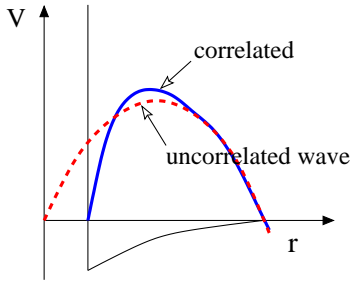}
\end{center}
\caption{Effect of the two-body correlations on the 
two-nucleon wave function as a function of the relative distance $r$. 
The nucleon-nucleon potential is schematically indicated. It shows the 
typical short-range repulsion ( hard core) together with the intermediate and 
long-range attractive parts.  
The figure is taken from \protect\cite{muether00}.}
\label{fig1}
\end{figure}

However, also within hadronic theories a perturbative approach to 
the strongly interacting nuclear systems 
is not possible. A systematic summation of diagrams 
up to infinite order in terms of the Brueckner hole-line expansion 
turned out to be an appropriate treatment. Already in lowest order which 
corresponds to standard Brueckner theory the saturation of nuclear 
matter can be described at least qualitatively \cite{gammel,coester70,liege}. 
In Brueckner theory the T-matrix (or Brueckner G-matrix) serves  
as an effective in-medium two-body interaction. It is
determined by a self-consistent summation of the ladder diagrams in a 
quasi-potential approximation (Thompson equation) to the Bethe-Salpeter 
equation. The character of the bare nucleon-nucleon interaction, in 
particular the repulsive short range part (hard-core) requires to account 
for two-body correlations in a self-consistent way. The effect of the 
correlations on the two-nucleon wave function in the medium is 
schematically depicted in Fig. \ref{fig1}.

However, non-relativistic Brueckner calculations are not able to 
meet the empirical saturation point of nuclear matter 
($\rho_{\rm sat} = 0.16~{\rm fm}^{-3}, ~E_{\rm bind} = -16$ MeV).  
In contrast, the saturation points obtained for 
various types of NN-potentials were all located on the so-called 
Coester line \cite{coester70} (for a recent review see \cite{muether00}) 
in the $\epsilon-\varrho$ plane. A breakthrough was achieved 
when first relativistic (Dirac-) Brueckner-Hartree-Fock (DBHF) 
calculations were 
performed in the eighties \cite{shakin,terhaar87,brockmann90}. Now 
the Coester line was shifted much closer towards the empirical area 
of saturation. One reason for the success of the relativistic approach 
is usually attributed to the fact that the dressing of the in-medium spinors 
introduces a density dependence to the interaction which is missing in the 
non-relativistic treatment. In the latter case the inclusion of 
three-body forces can lead to similar effects. The
occurrence of many-body forces is, however, closely connected to the 
inclusion of non-nucleon degrees of freedom, i.e. resonances. A
discussion of these aspects as well as a comparison of 
relativistic versus non-relativistic approaches will be given in
Sec. 5.  

Relativistic Brueckner calculations are not straightforward and the 
approaches of various groups \cite{terhaar87,brockmann90,weigel88,sehn97,fuchs98,boelting99,amorin92,dejong98} 
are similar but differ in detail, 
depending on solution techniques and the particular approximations made.  
The intention of the present work is to review the {\it standard} 
relativistic Brueckner approach and to discuss implications of the 
several approximation schemes, as well as the role of higher order
correlations, Pauli effects, and the special role of Dirac
phenomenology. 

Large part of the present discussion will be 
devoted to the determination of the nuclear self-energy. 
To determine its Lorentz structure and momentum dependence, 
the T-matrix has to be decomposed into Lorentz components, i.e. 
scalar, vector, tensor, etc. contributions. This procedure is not free 
from ambiguities \cite{nupp89}. Due to identical matrix elements for positive 
energy states pseudo-scalar and pseudo-vector 
components cannot uniquely be disentangled for on-shell scattering. 
However, with a pseudo-scalar vertex the pion couples 
maximally to negative energy 
states which are not included in the standard Brueckner approach. This 
is inconsistent with the potentials used since OBEPs are usually based 
on the no-sea approximation. Hence, pseudo-scalar contributions 
due to the one-$\pi$ exchange (OPE) lead to large and 
spurious contributions from negative energy states. In 
\cite{fuchs98} it was shown that such spurious contributions 
dominate the momentum dependence of the nuclear 
self-energy, and, in particular, lead to an artificially 
strong momentum dependence 
inside the Fermi sea. It was further demonstrated \cite{fuchs98} 
that previous methods \cite{terhaar87,sehn97} used 
to cure this problem fail and finally a new and reliable 
method was proposed to remove those spurious contributions 
from the T-matrix \cite{boelting99}. If calculations are 
performed in full Dirac space,
i.e. including anti-particles, the complete information on the Lorentz 
structure of the self-energy is available but in this case on has to
cope with other problems \cite{weigel88,amorin92,huber95,dejong98}.

\section{The Relativistic Brueckner Approach}
\subsection{The Coupled Set of Equations}
In the relativistic Brueckner approach the nucleon 
inside the nuclear medium is viewed as a dressed particle in consequence
of its two-body interaction with the surrounding nucleons. 
The in-medium interaction of the nucleons is treated in the ladder
approximation of the relativistic Bethe-Salpeter (BS) equation
\beq
T= V + i \int  VQGGT
\quad ,
\label{BSeq}
\eeq
where $T$ denotes the T-matrix. $V$ is the bare nucleon-nucleon interaction. 
The intermediate off-shell nucleons in the 
scattering equation are described by a two-particle propagator $iGG$.
The Pauli operator $Q$ accounts for the 
influence of the medium by the Pauli-principle and projects the 
intermediate scattering states out of the Fermi sea. 
The Green's function $G$ fulfills the Dyson equation
\beq
G=G_0+G_0\Sigma G 
\quad .
\label{Dysoneq}
\eeq 
$G_0$ denotes the free nucleon propagator while the influence of the 
surrounding nucleons is expressed by the nucleon self-energy $\Sigma$. 
In Brueckner theory this self-energy is determined by summing up the 
interaction with all the nucleons inside the Fermi sea in Hartree-Fock 
approximation
\beq
\Sigma = -i \int\limits_{F} (Tr[G T] - GT )
\quad .
\label{HFselfeq1}
\eeq
The coupled set of equations (\ref{BSeq})-(\ref{HFselfeq1}) represents
a self-consistency problem and has to be
iterated until convergence is reached.

Due to translational and rotational invariance, parity conservation 
and time reversal invariance the self-energy in isospin saturated 
nuclear matter has the general form 
$\Sigma = \Sigma_s - \gamma_\mu \Sigma^\mu$. It depends on the Lorentz 
invariants $k^2$, $k\cdot j$ and $j^2$, with $j_\mu$ and $k_\mu$ being the
baryon current and the nucleon four-momentum, respectively \cite{serot86}. 
The invariants can also be expressed in terms of 
$k_0, |{\bf k}|$ and $\kf$, where $\kf$ denotes the Fermi momentum.
Furthermore the vector part of the self 
energy has contributions proportional to $k^\mu$ and to the current 
$j^\mu$. Defining the streaming velocity as 
$u^\mu = j^\mu / \sqrt{j^2}$, the momentum $k^\mu$ can be decomposed 
into contributions parallel and perpendicular to the streaming velocity, i.e. 
$ k^\mu = (k\cdot u) u^\mu + \Delta^{\mu\nu}k_\nu$ with the 
projector $\Delta^{\mu\nu} = g^{\mu\nu} - u^\mu u^\nu$. 
The vector part of the self-energy can then be written covariantly 
as \cite{sehn97,amorin92} 
\begin{equation}
\Sigma^\mu = \Sigo u^\mu + \Sigv  \Delta^{\mu\nu}k_\nu
\quad . 
\label{sigvec1}
\end{equation}
Thus the full self-energy reads 
\begin{eqnarray}
\Sigma (k,\kf) &=& \Sigs (k,\kf) -\gamma_\mu \left[ \Sigo (k,\kf) \, u^\mu 
+ \Sigv (k,\kf) \, \Delta^{\mu\nu}k_\nu \right]
\label{sig1} \\
&=& \Sigs (k,\kf) -\gamma_0 \, \Sigo (k,\kf) + 
{\bf \gamma}\cdot {\bf k} \,\Sigv (k,\kf)\, |_{{\mathrm RF}} 
\label{sig2}
\end{eqnarray}    
where the subscript RF indicates the respective 
expressions in the nuclear matter rest frame 
($u^{\mu} = \delta^{\mu 0}$) \cite{terhaar87,horowitz87}. 
The $\Sigs,\Sigo$ and $\Sigv $ components are 
Lorentz scalar functions which actually 
depend on $k_0$,$|{\bf k}|$ and $\kf$. They follow from 
the self-energy matrix by taking the respective traces 
\cite{sehn97}
\begin{eqnarray}
\Sigs &=& \frac{1}{4} tr \left[ \Sigma \right] 
\label{trace1}\\
\Sigo &=& \frac{-1}{4} tr \left[ \gamma_\mu u^\mu \Sigma \right] 
       =  \frac{-1}{4} tr \left[ \gamma_0 \, \Sigma \right]_{{\mathrm RF}}
\label{trace2}\\
\Sigv &=& \frac{-1}{4\Delta^{\mu\nu}k_\mu k_\nu } 
          tr \left[\Delta^{\mu\nu}\gamma_\mu k_\nu \, \Sigma \right] 
       = \frac{-1}{4|{\bf k}|^2 } 
tr \left[{\bf \gamma}\cdot {\bf k} \, \Sigma \right]_{{\mathrm RF}} 
\label{trace3}
\quad .
\end{eqnarray}

The Dirac equation for the in-medium spinor basis can be deduced from the 
Green's function. Written in terms of effective masses and momenta 
\begin{eqnarray}
\quad m^* = M+ Re \, \Sigs \quad , \quad k^{*}_\mu = k_\mu + Re \, \Sigma_\mu 
\label{sig3}
\end{eqnarray}
the Dirac equation has the form
\begin{equation}
\left[ k^* \!\!\!\!\!\! / - m^* -i \, Im \, \Sigma \right] u(k) =0 .
\label{dirac} 
\end{equation}
In the following we will work in the quasi-particle approximation 
and neglect the imaginary part of the
self-energy from now on. Thus the effective nucleon four-momentum will be 
on mass shell even above the Fermi surface, fulfilling the relation 
$ k^{*}_\mu k^{*\mu} = m^{* 2}$. 
Since we only deal with the real part 
of the self-energy in the quasi-particle approximation we omit this in the 
notation. In the nuclear matter rest frame the four-momentum 
follows from Eq. (\ref{sig3})
\begin{eqnarray}
{\bf k}^* = {\bf k} (1+\Sigv) 
\quad , \quad
k^{*}_0 = E^* = \sqrt{ {\bf k}^2 (1+\Sigv)^2 + m^{*2} }
\label{estar}
\end{eqnarray}
which allows one to eliminate the $\Sigv$-term in the Dirac equation, 
\begin{eqnarray}
\left[ ({\bf \alpha} \cdot {\bf k}) - \gamma^0  \mst \right] u(k) =\est u(k) 
\quad ,\label{dirac2}
\end{eqnarray} 
by a rescaling of the effective mass and 
the kinetic energy
\begin{eqnarray} 
\mst = \frac{m^*}{1+\Sigv} \quad , \quad \est = \frac{E^*}{1+\Sigv} 
     = \sqrt{{\bf k}^2 + \mstq }
\quad . 
\label{red1}
\end{eqnarray}
The solution of the Dirac equation provide the in-medium nucleon
spinor basis
\beq
u_\lambda (k,\kf)= \sqrt{ { {\tilde E}^*({\bf k})+ {\tilde m}^*_F}\over
{2{\tilde m}^*_F}} 
\left( 
\begin{array}{c} 1 \\ 
{2\lambda |{\bf k}|}\over{{\tilde E}^*({\bf k})+ {\tilde m}^*_F}
\end{array}
\right)
\chi_\lambda
\quad ,
\label{spinor}
\eeq
where ${\tilde E}^*({\bf k})=\sqrt{{\bf k}^2+{\tilde m}^{*2}_F}$.
$\chi_\lambda$ denotes a two-component Pauli spinor with 
$\lambda=\pm {1\over 2}$. The normalization of the Dirac spinor is 
$\bar{u}_\lambda(k,\kf) u_\lambda(k,\kf)=1$. 
Since the in-medium spinor contains the reduced effective 
mass the matrix elements of the 
bare nucleon-nucleon interaction become density dependent. 
From the Dirac equation (\ref{dirac2})
one derives the relativistic Hamiltonian, i.e. the single-particle potential 
${\hat U } = \gamma^0  \Sigma$. 
The expectation value of ${\hat U}$, i.e. sandwiching ${\hat U}$ between 
the effective spinor basis (\ref{spinor}), yields the single particle potential 
\begin{eqnarray}
   U(k) = \frac{\langle u(k)|\gamma^0  \Sigma | u(k)\rangle }
{\langle u(k)| u(k)\rangle } =
\frac{m^{\ast}}{E^{\ast}({\bf k})} 
\, \langle {\bar u(k)}| \Sigma | u(k)\rangle
\label{upot1}
\end{eqnarray}
which can be evaluated as 
\begin{eqnarray}
U(k,\kf) &=& \frac{m^*}{E^*} \Sigs - \frac{ k_{\mu}^* \Sigma^\mu}{E^*} \\
         &=& \frac{m^* \Sigs }{\sqrt{ {\bf k}^2 (1+\Sigv)^2 + m^{*2}}} 
         - \Sigo + \frac{ (1+\Sigv)\Sigv {\bf k}^2}
           {\sqrt{ {\bf k}^2 (1+\Sigv)^2 + m^{*2}}}
\quad .
\label{upot2}
\end{eqnarray}
In many applications \cite{brockmann90,lee97} the single particle 
potential is only given in terms of a scalar and zero-vector component. 
This can be achieved by introducing reduced fields ${\tilde \Sigs}$ and 
${\tilde \Sigo}$ as 
\begin{eqnarray}
{\tilde \Sigs} = \mst -M = \frac{\Sigs - \Sigv M}{1+\Sigv} 
\quad , \quad 
{\tilde \Sigo} = \est -E = \Sigo - \est ({\bf k}) \Sigv
\quad . 
\label{red2}
\end{eqnarray}
The single particle potential has then the form
\begin{equation}
U(k,\kf) = \frac{\mst}{\est} {\tilde \Sigs} - {\tilde \Sigo} 
\quad . 
\label{upot3}
\end{equation}
Frequently the reduced fields, Eq. (\ref{red2}), are 
used rather than the projected components since they represent 
the self-energy in a mean field or Hartree form. Thus they can 
easily be related to effective hadron mean field theory 
\cite{fuchs95,toki97}. Such a representation is 
meaningful since the $\Sigv$-contribution is a moderate 
correction. 
\subsection{The In-Medium T-Matrix}
Before going into details I will shortly summarize the 
main assumptions which are made in the {\it standard} relativistic Brueckner 
approach to solve the BS-equation (\ref{BSeq}): 
\begin{itemize}
\item {\it No sea approximation}. The subspace of negative energy 
states is omitted. In this way one avoids the delicate problem 
of infinities which would generally appear due to 
contributions from negative energy nucleons in the Dirac sea. 
The approximation is consistent with the usage of standard OBE
potentials which are derived under the same assumption.\\ 

\item {\it Thompson choice}. The full two-body propagator $iGG$ in
the BS-equation is replaced by an effective two-body propagator 
propagator. The Thompson propagator (and similar the 
Blankenbecler sugar propagator) projects the intermediate nucleons 
onto positive energy states and restricts the exchanged energy
transfer by $\delta(k^0)$ to zero. Thus  
the BS-equation is reduced to a three dimensional integral equation of
the Lippmann-Schwinger type, the so called Thompson equation 
\cite{thompson70}. \\

\item {\it Reference spectrums approximation}. The momentum dependent 
effective mass $\mst$ which enters into the Thompson propagator is 
replaced by an average value ${\tilde m}^*_F$ (averaged over the Fermi sea). The
approximation is justified as long as the self-energy exhibits a weak 
momentum dependence.\\

\item {\it Angle-averaged Pauli operator}. The Pauli operator is
replaced by its angle-averaged counterpart which allows to solve 
the Thompson equation in a decoupled angular-momentum partial wave basis. \\

\item {\it Quasi-particle approximation}. The T-matrix is determined 
for on-shell scattering at the quasi-particle pole. Finite width
spectral functions are not taken into account.

\end{itemize}
In contrast to the self-energy, Eq. (\ref{HFselfeq1}), 
which has to be calculated in the nuclear matter rest frame,  
the Thompson Eq. (\ref{thompsoneq}) is most naturally solved in 
the two-nucleon c.m. frame. 
The Thompson propagator and similar the Blankenbecler-Sugar propagator 
imply that the time-like component of the momentum transfer in $V$ and $T$ 
is set equal to zero which is a natural constraint in the c.m. frame,
however, not a covariant one. The Thompson equation reads in the c.m. frame
\beqa
T({\bf p},{\bf q},x)|_{c.m.} &=&  V({\bf p},{\bf q}) 
\label{Tmateq}\\
&+& 
\int {d^3{\bf k}\over {(2\pi)^3}}
{\rm V}({\bf p},{\bf k})
{{\tilde m}^{*2}_F\over{\tilde{E}^{*2}({\bf k})}}
{{Q({\bf k},x)}\over{2{\tilde{E}}^*({\bf q})-2{\tilde{E}}^*({\bf k})
+i\epsilon}}
T({\bf k},{\bf q},x) 
\nonumber
\quad ,
\label{thompsoneq}
\eeqa
where ${\bf q}=({\bf q}_1 - {\bf q}_2)/2$ is the relative three-momentum 
of the initial state while $\bf k$ and $\bf p$ are the relative 
three-momenta of the intermediate and final states, respectively. 
The starting energy in Eq. (\ref{thompsoneq}) is already fixed by 
$\sqrt{\tilde{s}^*}=2{\tilde{E}}^*({\bf q})
=2\sqrt{{\bf q}^2+\tilde{m}^{*2}_F}$. If ${\bf q}_1$ and ${\bf q}_2$ are 
nuclear matter rest frame momenta of the nucleons in the initial state, 
the boost-velocity $\bf u$ into the c.m. frame is given by 
\beq
{\bf u} = {\bf P}/ \sqrt{\tilde{s}^{*}+{\bf P}^2}
\label{boost}
\quad ,
\eeq
with the total three-momentum and the invariant mass 
${\bf P} = {\bf q}_1+{\bf q}_2$ and $\tilde{s}^*=(\tilde{E}^*({\bf q}_1)
+\tilde{E}^*({\bf q}_2))^2-{\bf P}^2$, respectively.
In Eq. (\ref{thompsoneq}) $x$ denotes the set of 
additional parameters $x=\{\kf, {\tilde m}^*_F,|{\bf u}|\}$ 
on which the T-matrix depends. 

The Pauli operator $Q$ explicitely 
depends on the chosen frame, i.e., on  the 
boost 3-velocity ${\bf u}$ into the c.m.-frame. 
The Thompson equation (\ref{thompsoneq}) for the on-shell T-matrix
$(|{\bf p}|=|{\bf q}|)$ can be solved applying standard 
techniques described in detail by Erkelenz \cite{erkelenz74}. 
Doing so, one constructs the positive-energy helicity T-matrix elements from the 
$|JMLS\rangle$-scheme. On-shell only five of the sixteen helicity 
matrix elements are independent which follows from general 
symmetries \cite{erkelenz74}. After a partial wave 
projection onto the $|JMLS\rangle$-states the integral reduces to a one-dimensional 
integral over the relative momentum $|{\bf k}|$ and Eq. (\ref{thompsoneq}) decouples
into three subsystems of integral equations for the uncoupled spin singlet,
the uncoupled spin triplet and the coupled triplet states.
For this purpose the Pauli operator $Q$ has to be replaced 
by an angle averaged Pauli operator
$\overline Q$ \cite{horowitz87}. We are solving the integral equations by the
matrix inversion techniques of Haftel and Tabakin \cite{haftel70}. 
Real and imaginary parts of the T-matrix are calculated separately 
by the principal-value treatment given in  Ref. \cite{trefz85}. 
Due to the anti-symmetry of the two-fermion states the total
isospin I of the two-nucleon system (I=0,1) can be restored 
by the selection rule: 
\beq
(-)^{L+S+{\rm I}}=-1~~.
\label{srule}
\eeq
From the five independent on-shell amplitudes in the $|JMLS\rangle$-representation
the five independent partial wave amplitudes in the helicity representation
(for I=0,1 and real and imaginary part separately) are obtained
by inversion of Eq. (3.32) and then of Eq. (3.28) of Ref. \cite{erkelenz74}.
The summation over the total angular momentum $J$ 
yields the full helicity matrix element
\begin{equation}                       \label{eq12}
\sum_J \left[ \frac{2J+1}{4\pi} \right]
d^J_{\lambda \lambda^{\prime} }(\theta)
\langle|{\bf p}|\lambda^{\prime}_1 \lambda^{\prime}_2 |T^{J,{\rm I}}(x)|
\, |{\bf q}| \lambda_1 \lambda_2\rangle
= \langle{\bf p} \lambda^{\prime}_1 \lambda^{\prime}_2 {\rm I} {\rm I}_3 | T(x)
| {\bf q} \lambda_1 \lambda_2 {\rm I} {\rm I}_3 \rangle .
\label{helampl}
\end{equation}
Here $\theta$ is the scattering angle between ${\bf q}$ and ${\bf p}$ and 
$\lambda = \lambda_1 - \lambda_2, \lambda' = \lambda_1' - \lambda_2'$.
The reduced rotation matrices $d^J_{\lambda \lambda^{\prime} }(\theta)$ 
are those defined by Rose \cite{rose}.
The matrix element (\ref{eq12}) is actually independent of the third
component of the isospin ${\rm I}_3$.
\section{The Nuclear Self-Energy}
The easiest way to determine scalar and vector self-energy 
components directly from the single particle potential $U$
(\ref{upot1}). Since $U$ is obtained after complete summation over 
Dirac-indices of the T-matrix one by this way can avoid 
cumbersome projection techniques which are required using the trace 
formulas (\ref{trace1})-(\ref{trace3}). 
Using Eq. (\ref{upot3}) a fit to $U$ delivers the 
density dependent but  {\it momentum independent } self-energy 
components ${\tilde \Sigs}$ and ${\tilde \Sigo}$. This method has
e.g. been applied in  \cite{brockmann90}. An attempt to extend 
this method  and to extract by fitting procedures  momentum 
dependent fields \cite{lee97} suffered by large uncertainties 
since one tries then to extract two functions out of one.

A more accurate determination of the density and momentum 
dependence of the self-energy requires projection techniques for 
the in-medium T-matrix as outlined first by Horowitz and Serot 
\cite{horowitz87}. That this procedure is also not free from ambiguities has 
been noticed relatively early \cite{terhaar87,nupp89}. 
The whole problem arises from the {\it no sea approximation} 
in the standard approach. When calculations are performed in full
Dirac space \cite{weigel88,amorin92,dejong98,huber95} the Lorentz
structure of the self-energy can uniquely determined from the
information available form those matrix elements 
($ \langle \bar{v}|\Sigma | u\rangle ,\langle \bar{v}|\Sigma | v\rangle $) 
which involve negative energy states. The inclusion of negative energy 
excitations with 4 states for each spinor yields in total 
$4^4=256$ two-body matrix elements for the T-matrix. 
Symmetry arguments reduce this to 44 for on-shell
particles \cite{tjon85b}. If one takes now only positive energy solutions into account
this reduces to $2^4=16$ two-body matrix elements. 

For on-shell matrix elements the number of independent matrix elements can be further
reduced by symmetry arguments down to 5. Thus, all on-shell two-body matrix elements
can be expanded into five Lorentz invariants. But these five invariants are not 
uniquely determined since the Dirac matrices involve also negative energy states. 
The decomposition of a one-body operator into a Lorentz scalar and
a Lorentz vector contributions depends therefore on the choice of 
these five Lorentz invariants. 

In nuclear matter the largest ambiguity arises concerning 
the determination of pseudo-scalar ($ps$) and pseudo-vector ($pv$) 
T-matrix elements. The $pv$ invariant in the medium is defined as
\beq
PV =  \frac{k_{2}^* \!\!\!\!\! / - k_{1}^* \!\!\!\!\! /}{2m^{\ast}}\gamma_5
\otimes 
\frac{q_{2}^* \!\!\!\!\! / - q_{1}^* \!\!\!\!\! /}{2m^{\ast}}\gamma_5
\eeq
with $k_{1}^*,q_{1}^* $ the initial and $k_{2}^*,q_{2}^* $ 
the final momenta of the scattering particles. For on-shell 
scattering of positive energy states the $ps$ and $pv$ matrix elements 
are identical (using the Dirac eq.) 
\beq 
{\overline u}(q )\left( 
\frac{q^* \!\!\!\!\! / - p^* \!\!\!\!\! /}{2m^{\ast}}  
\right) \gamma_5 u(p ) = {\overline u}(q )\gamma_5 u(p )
\quad .
\label{mat1}
\eeq
The $ps$ vertex couples on the other hand maximally to negative energy 
whereas the $pv$ vertex suppresses the coupling 
to antiparticles for on-shell scattering
\beq 
{\overline v}(q )\left(   
\frac{q^* \!\!\!\!\! / - p^* \!\!\!\!\! /}{2m^{\ast}} 
\right) \gamma_5 u(p ) = 0 
\quad .
\label{mat2}
\eeq
To summarize: The $ps$ and $pv$ matrix elements of an on-shell 
two-body operator, e.g. the T-matrix, can (in the positive energy 
sector) not be determined uniquely by projection techniques. If the 
matrix elements are known {\it a priori}, as e.g. for the bare 
NN interaction $V$, of course no problems arise. The same holds when 
the Dirac sea is included in the formalism. The full information on 
the T-matrix is available and the Lorentz structure of the self-energy 
is then uniquely determined \cite{weigel88,amorin92,dejong98}. 
However, such an approach suffers from other problems (see next
section). 
\begin{figure}
\begin{center}
\leavevmode
\includegraphics[width=.8\textwidth]{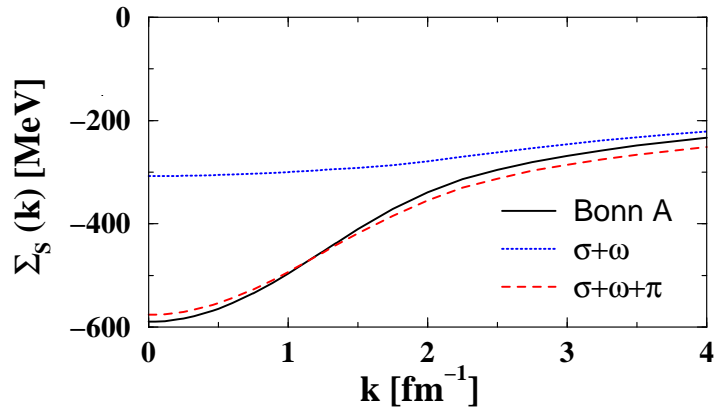}
\end{center}
\caption{
Influence of the various meson exchange contributions on the 
nucleon self-energy (scalar part). The solid line corresponds to 
the full calculation (Bonn A), the dotted line to $\sigma\omega$ and 
the dashed line to $\sigma\omega\pi$--exchange only. In all 
calculations the {\it pv choice} is used. 
}
\label{sig1_fig}
\end{figure}
The ambiguity problem has e.g. been pointed out by Nuppenau 
et al. \cite{nupp89} and ter Haar and Malfliet \cite{terhaar87} 
proposed a recipe to cure which was used by various groups:    
perform the projection, take the  $ps$ matrix element and 
replace it (due to physical reasons) by a $pv$ one. This procedure 
was called the {\it pseudo-vector choice}. Later on, 
this procedure was critically examined by Fuchs et al. \cite{fuchs98} 
and it was shown that it completely fails in controlling 
the leading order $1-\pi$-exchange contribution. 

Fig. \ref{sig1_fig} shows the momentum dependence 
of the scalar self-energy component $\Sigs$ at  
nuclear matter density $\rho = 0.166\,{\rm fm}^{-3}$ obtained in  
the {\it pv choice} projection scheme, as it arises from the 
various meson exchange contributions of the Bonn A potential. 
Taking only $\sigma$ and $\omega$ exchange into account the 
the momentum dependence is flat inside the Fermi sea. Including 
the pion we are already very close to the full DBHF result. The 
strong momentum dependence of the present calculation originates to 
a large extent from pion-exchange. 

\begin{figure}
\begin{center}
\leavevmode
\includegraphics[width=.8\textwidth]{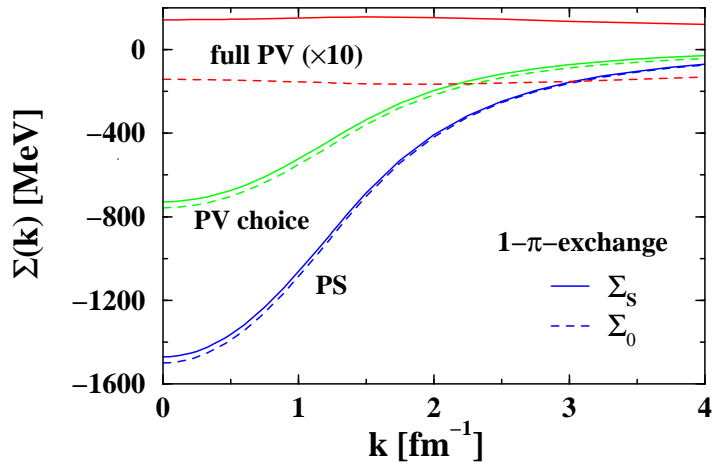}
\end{center}
\caption{Hartree-Fock self-energy originating from OPEP. 
results for $V_\pi$ with $ps$ and $pv$ coupling are 
compared to results obtained within the {\it pv choice}. 
Solid lines represent the scalar, 
dashed lines the vector self-energy. 
}
\label{sig_pi_fig}
\end{figure}

It is a well known fact that a pseudo-scalar $\pi NN$ coupling 
leads to extremely large pion contributions to the nuclear self-energy 
and contradicts soft pion theorems of ChPT \cite{chpt}. Therefore 
in OBEPs always $V_\pi$ with  pseudo-vector $\pi NN$ coupling is
used. It is, 
however, instructive to test the {\it pv choice} projection recipe 
for the case of the $\pi$-exchange \cite{fuchs98}. 
This is done in Fig. \ref{sig_pi_fig} where the Hartree-Fock 
self-energy from the 1-$\pi$-exchange potential  
(OPEP) is shown. Exact results 
for a $ps$ and $pv$ $\pi$NN coupling can be compared to those 
obtained by projection techniques. It is seen that a $ps$ 
description OPEP leads to extremely large self-energy 
components and a very strong momentum dependence. The $pv$ 
coupling suppresses the $V_\pi$ contribution 
by nearly two orders of magnitude and even on that scale 
the momentum dependence is much less 
pronounced. At the Hartree-Fock level the $V_\pi$ self-energy 
can be computed directly or, alternatively, applying the same 
techniques as for the full T-matrix, i.e. going through the
transformations from the $|LSJ\rangle$ basis to the helicity basis and
finally via projection to the basis of covariant amplitudes.  
Doing so,  it turns out that the {\it pv choice} projection 
fails to describe a $pv$ pion 
exchange. Thus it is clear that the strong momentum dependence 
seen in the full self-energy (Fig. \ref{sig1_fig}) is to large 
part due to spurious contributions form pseudo-scalar $\pi$-exchange. 
How these contributions can be eliminated has been discussed in 
\cite{fuchs98} and \cite{boelting99} and is explained in more 
detail below.

\subsection{Covariant Representation of the T-Matrix}
To use the trace formulas, Eqs. (\ref{trace1}-\ref{trace3}), one has to represent the 
T-matrix covariantly. A set of five linearly independent covariants is 
sufficient because on-shell only 
five helicity matrix elements appear as solution of the Thompson equation. 
A linearly independent although not unique set of five covariants 
is given by the Dirac covariants 
\beqa
\!\!\!\!\!{\rm S} = 1\otimes 1 ,
{\rm V} =  \gamma^{\mu}\otimes \gamma_{\mu},
{\rm T} = \sigma^{\mu\nu}\otimes\sigma_{\mu\nu}, 
{\rm A} =  \gamma_5 \gamma^{\mu}\otimes \gamma_5 \gamma_{\mu},
{\rm P} = \gamma_5 \otimes \gamma_5 .
\label{inv}
\eeqa
Using this special set, dubbed as $ps$ representation in 
the following, the on-shell T-matrix for definite isospin 
I can be represented covariantly as \cite{horowitz87}
\beqa
\hspace{1cm}
T^{\rm I}(\pabs,\theta,x)&=& 
 F_{\rm S}^{\rm I}(\pabs,\theta,x){\rm S}
+F_{\rm V}^{\rm I}(\pabs,\theta,x){\rm V}
+F_{\rm T}^{\rm I}(\pabs,\theta,x){\rm T}
\nonumber \\
&+&F_{\rm A}^{\rm I}(\pabs,\theta,x){\rm A}
+F_{\rm P}^{\rm I}(\pabs,\theta,x){\rm P}
\label{tmatrep1}
\quad .
\eeqa
Here ${\bf p}$ and $\theta$ denote the relative three-momentum and the
scattering angle between the scattered nucleons in the c.m. frame, 
respectively. The {\it direct} (Hartree) amplitudes are are given by 
$\theta =0$ and the {\it exchange} amplitudes (Fock) by $\theta
=\pi$. The five covariant amplitudes $F_{\rm i}^{\rm I}$ can 
be obtained by matrix inversion of eq. (\ref{tmatrep1}) from the 
five helicity amplitudes $T^{\rm I}$. The nucleon self-energy 
in isospin saturated nuclear matter 
has then the form \cite{sehn97}
\beq
\Sigma_{\alpha\beta}(k,\kf)= \int {{d^3{\bf q}}\over {(2\pi)^3}}
{{\theta(\kf-|{\bf q}|)}\over {\tilde{E}^*({\bf q})}}
\left[\tilde{m}^*_F 1_{\alpha\beta}F_{\rm S}
+\not{\tilde q}^*_{\alpha\beta}F_{\rm V}\right]
\quad ,
\label{self1}
\eeq
where the isospin averaged amplitudes are defined as
\beq
F_i(|{\bf p}|,0,x):=
{1\over 2}\left[ F_i^{{\rm I}=0}(|{\bf p}|,0,x)
+3 F_i^{{\rm I}=1}(|{\bf p}|,0,x)\right]
\quad .
\eeq
Eq. (\ref{self1}) shows that the self-energy can be expressed 
solely in terms of direct scalar 
and vector amplitudes $F_{S,V}$, if these are derived from 
already anti-symmetrized helicity amplitudes (\ref{helampl}) 
which obey the selection rule (\ref{srule}).  
Corresponding covariant exchange amplitudes 
$F_{\rm i}^{\rm I}(\pabs,\pi,x)$ are obtained from the 
exchange helicity amplitudes inverting a similar matrix for 
the  exchange invariants $\tilde{\rm S},\tilde{\rm V},
\tilde{\rm T},\tilde{\rm A},\tilde{\rm P}$. The latter are 
related to the original invariants (\ref{inv}) through a Fierz 
transformation. The same Fierz transformation relates also 
direct and exchange amplitudes $F_{\rm i}$. Thus the 
direct scalar and vector amplitudes $F_{S,V}$ in Eq. (\ref{self1}) 
contain already contributions from all other exchange 
amplitudes \cite{tjon85b,fuchs98}. 

Hence an explicit splitting 
of the already anti-symmetrized helicity amplitudes into 
direct and exchange parts 
$\frac{1}{2}(T^{\rm I}(\pabs,0,x) - T^{\rm I}(\pabs,\pi,x))$ 
provides no additional information and is also not necessary. 
It becomes only relevant if one wants to replace the $ps$ 
invariant by the $pv$ invariant. However, now it becomes 
evident why the the above mentioned {\it pv choice} 
does not lead to the desired result: In this procedure the 
pseudo-scalar exchange amplitude $F_{\rm P}^{\rm I}(\pabs,\pi,x)$
is kept fixed and interpreted as a pseudo-vector one, 
replacing the corresponding invariant 
$\tilde{\rm P}\longmapsto \widetilde{\rm PV}$. Since the other 
amplitudes, due to Fierz, contain also $ps$ contributions such 
an replacement is incomplete. As can be seen from
Fig. \ref{sig_pi_fig} the spurious $ps$ contributions of the 
other amplitudes are still large. 

To eliminate the such spurious pseudo-scalar contributions of 
the 1-$\pi$ exchange completely one has to switch to another 
covariant representation of the T-matrix proposed by Tjon and Wallace 
\cite{tjon85a} which we call {\it full pv} representation in the following
\beqa
\hspace{1cm}
T^{\rm I}(\pabs,\theta,x)&=& g_{\rm S}^{\rm I}(\pabs,\theta,x){\rm S}
-g_{\tilde{\rm S}}^{\rm I}(\pabs,\theta,x)\tilde{\rm S}
+g_{\rm A}^{\rm I}(\pabs,\theta,x)({\rm A}-\tilde{\rm A})\nonumber \\
&+& g_{\rm PV}^{\rm I}(\pabs,\theta,x){\rm PV}
-g_{\widetilde{\rm PV}}^{\rm I}(\pabs,\theta,x)\widetilde{\rm PV}
\quad .
\label{tmatrep6}
\eeqa
The amplitudes $g^{\rm I}(\theta)$ are explicitly given in
\cite{boelting99}. In this scheme the pseudo-vector OPEP is exactly recovered.
\begin{figure}
\begin{center}
\leavevmode
\includegraphics[width=1.0\textwidth]{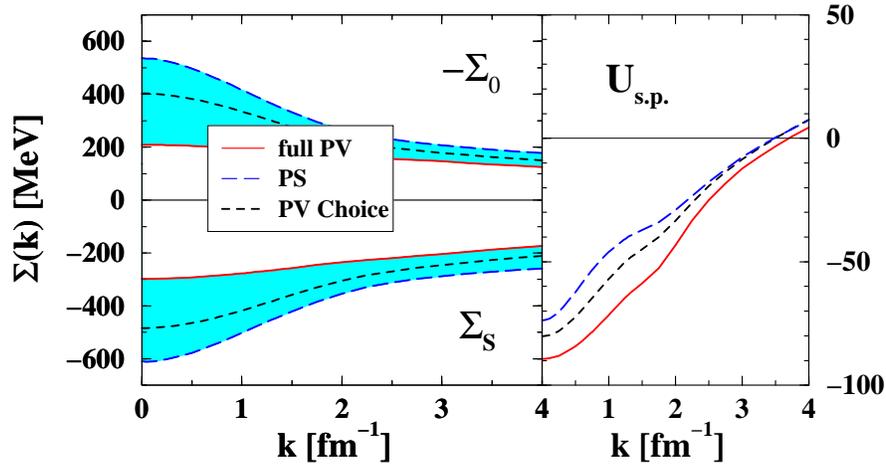}
\end{center}
\caption{
Range of uncertainty spanned by the various decomposition schemes of
the T-matrix for: self-energy components (left); single particle
potential  (right). The nuclear matter density is chosen as 
$\rho=0.166\,{\mathrm fm}^{-3}$ and the Bonn A potential is used.
}
\label{sig2_fig}
\end{figure}
In Fig. \ref{sig2_fig} the corresponding self-energies obtained 
for the various decompositions are compared. Adopting the 
{\it full pv} representation, the space-like $\Sigv$ contribution 
turns out to be much smaller than in the $ps$ or the 
standard {\it pv choice}. Therefore we show the reduced self-energies 
${\tilde \Sigs}$ and  ${\tilde \Sigo}$ in which $\Sigv$ 
is included for a better comparison.  
The pure $ps$ and the {\it full pv} representation can be regarded as the limiting 
cases which give the range of uncertainty in the determination 
of the self-energy. The latter has the big advantage 
that this method ensures by 
construction a correct treatment of the $pv$ OPEP 
at the Hartree-Fock level. Although the range of uncertainty is 
a few 100 MeV at the level of the self-energy components, it drops
out to most extent for {\it physical} observables which are based on 
complete matrix elements where vector and scalar parts 
contribute always with different sign. E.g., at 
the level of the single particle potential $U$ the remaining
uncertainty is only of about $10\div 20$ MeV.

\subsection{Covariant Representations of the Subtracted T-Matrix}
The {\it full pv} representation successfully 
reproduces the HF nucleon self-energy for the pion exchange with 
$pv$ coupling. However, as pointed out in \cite{fuchs98}, the {\it full pv} 
representation fails to reproduce the HF nucleon self-energy of other 
meson exchange potentials. Hence, it appears reasonable to treat 
the bare interaction $V$ and the higher order ladder graphs 
of the T-matrix separately. Since the OBEPs are known analytically 
we can use a mixed representation of the form 
\beq
V=V_{\pi,\eta}^{PV}+V_{\sigma,\omega,\rho,\delta}^{P}
\quad .
\label{mixed1}
\eeq
Here the $\pi$- and $\eta$-amplitudes are treated by the 
decomposition (\ref{tmatrep6}) while for the
$\sigma,\omega,\rho,\delta$-amplitudes  
the $ps$ representation (\ref{tmatrep1})  is applied. 
The higher order correlations of the T-matrix 
\beq
T_{Sub}=T-V=i\int VQGGT = \sum\limits_{n=1}^\infty \int V(iQGGV)^n
\quad ,
\eeq
in the following called the subtracted T-matrix, can not be represented 
by such a mixed form since one can not disentangle the different
meson contributions in the correlated ladder diagrams. 
The representation of the subtracted T-matrix remains therefore ambiguous.
However, if the pion exchange contributes dominantly at 
the Hartree-Fock level a $ps$ representation of the subtracted T-matrix 
should be more appropriate because of the higher order contributions of 
other meson exchanges. Thus the most favorable representation of 
the T-matrix is given by the $ps$ representation
\beq
T^{\rm P}=T^{\rm P}_{Sub}+V_{\pi,\eta}^{PV}+V_{\sigma,\omega,\rho,\delta}^{P}
\quad .
\label{tmatps}
\eeq
Here the $ps$ representation for $T^{\rm P}_{Sub}$ is determined via the 
matrix elements
\beqa
\langle{\bf p}\lambda_1^{'}\lambda_2^{'}|T^{\rm I}_{Sub}(x)|
{\bf q}\lambda_1\lambda_2\rangle:=
\langle{\bf p}\lambda_1^{'}\lambda_2^{'}|T^{\rm I}(x)-V^{\rm I}(x)|
{\bf q}\lambda_1\lambda_2\rangle 
\quad ,
\label{tmatsub}
\eeqa
with subsequently applying the projection scheme as in Eq. (\ref{tmatrep1}).
An alternative representation of the T-matrix is given
by a representation
\beq
T^{\rm PV}=T^{\rm PV}_{Sub}+V_{\pi,\eta}^{PV}+V_{\sigma,\omega,\rho,\delta}^{P}
\quad ,
\label{tmatpv}
\eeq
where the subtracted T-matrix is represented by the {\it full pv} 
representation (\ref{tmatrep6}). This representation is similar
to the {\it full pv} representation of the full T-matrix, however, with the
advantage that now the pseudo-scalar contributions in the bare 
nucleon-nucleon interaction, e.g. the 1-$\omega$ exchange potential, 
are represented correctly.

In \cite{boelting99} the two 
representations (\ref{tmatps}) and (\ref{tmatpv}) for the 
higher order ladder graphs were studied in detail. 
These two representations set the range of the remaining 
ambiguity concerning the representation of the T-matrix, i.e. after
separating the leading order contributions. The outcome is the 
following:
\begin{itemize}
\item The dependence of the ladder kernel on the two different 
representation schemes is generally weak. This gives confidence 
that the ambiguities are to most extent removed as long as the 
leading Born term, in particular the $pv$ OPEP, is treated 
correctly within the projection scheme.
\item The momentum dependence of the self-energy is moderate 
and close to the {\it full pv} case shown in Fig. \ref{sig2_fig}. 
This observation also justifies the {\it reference spectrums
approximation}. 
\end{itemize}
Larger differences between the $ps$ and $pv$ representations 
of the ladder kernel occur only at high densities. Here the 
$ps$ representation appears to be more appropriate. Therefore 
we consider this as the best solution of the problem. More 
details can be found in \cite{boelting99}. In this context it 
should be noted that in \cite{schiller} the present subtraction 
scheme was extended to a special treatment of $V_\rho$ where 
the tensor part leads also to a similar, but small on-shell 
ambiguity due to the Gordon identity.

\section{Nuclear Matter}
In this section I will discuss the role of correlations as well 
as implications of the various approximation schemes. Main emphasis 
will thereby be put on nuclear bulk properties, in particular on the 
nuclear equation-of-state (EOS), i.e. the binding energy
per particle, and the nuclear saturation mechanism.
\subsection{The Equation-of-State}
\begin{figure}
\begin{center}
\leavevmode
\includegraphics[width=.8\textwidth]{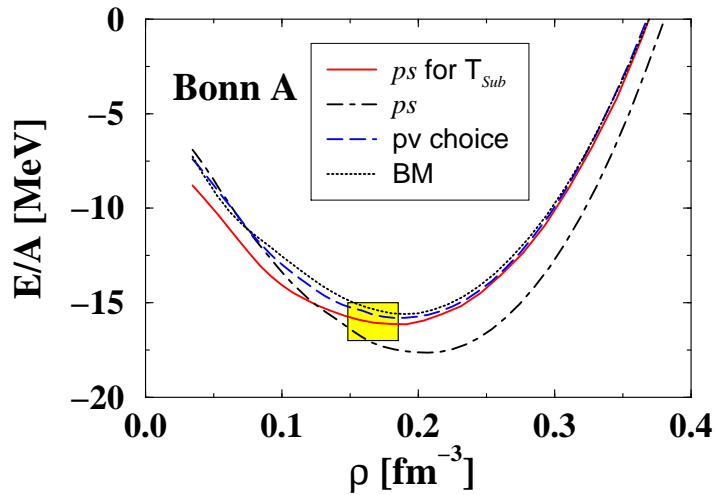}
\end{center}
\caption{Binding energy per particle as a 
function of nuclear matter density. 
As bare nucleon-nucleon interaction the Bonn A potential 
is used. For the T-matrix the subtraction scheme with the $ps$ representation 
for the ladder kernel (solid) is compared to a $ps$ representation of
the full T-matrix (dash-dotted), to the {\it pv choice} and to the 
result of \protect\cite{brockmann90} (BM, dotted).
}
\label{eos1_fig}
\end{figure}

In the relativistic Brueckner theory the energy per particle is defined 
as the kinetic plus half the potential energy 
\beq
E/A  =  {1\over \rho}\sum_{{\bf k},\lambda}
\langle\bar{u}_\lambda({\bf k})| 
\bfgamma \cdot {\bf k} + M + {1\over 2}\Sigma (k)
| u_\lambda({\bf k})\rangle
{{\tilde{m}^*(k)}\over {{\tilde{E}}^*(k)}} - M
\quad .
\label{eos}
\eeq
In Fig. \ref{eos1_fig} the EOSs, obtained in the various treatments, 
are compared. All calculations are based on the Bonn A interaction. 
First of all one sees that, except for a  
full $ps$ treatment which is not correct for realistic
potentials, the different calculations coincide at high
densities. If one applies the {\it pv choice} the result is very close 
to that obtained by Brockmann and Machleidt (BM) \cite{brockmann90}. 
This is somewhat surprising since there no 
projection scheme to the T-matrix has been applied but constant, i.e. 
momentum independent, self-energy components have been determined by 
a fit to the single particle potential.  
As discussed in the previous sections the 
{\it pv choice} fails to reproduce the $pv$ OPEP contribution 
to the self-energy. One can 
estimate this effect at the level of the binding energy 
by the comparison of the $ps$ representation for the subtracted and the 
$ps$ representation for the full T-matrix. 
In the latter case the nucleons are less bound at small 
densities. The situation changes, however, around saturation density. 
The full $ps$ representation of the T-matrix 
contains maximal contributions from a pseudo-scalar $\pi NN$ 
coupling which leads to saturation properties closer 
to non-relativistic Brueckner calculations (see below). 
A correct pseudo-vector representation
of the pion, as used in the subtraction scheme,  
suppresses this effect. Thus at smaller densities one obtains a larger
binding, while around saturation density the EOS is more repulsive. 

\begin{figure}
\begin{center}
\leavevmode
\includegraphics[width=1.0\textwidth]{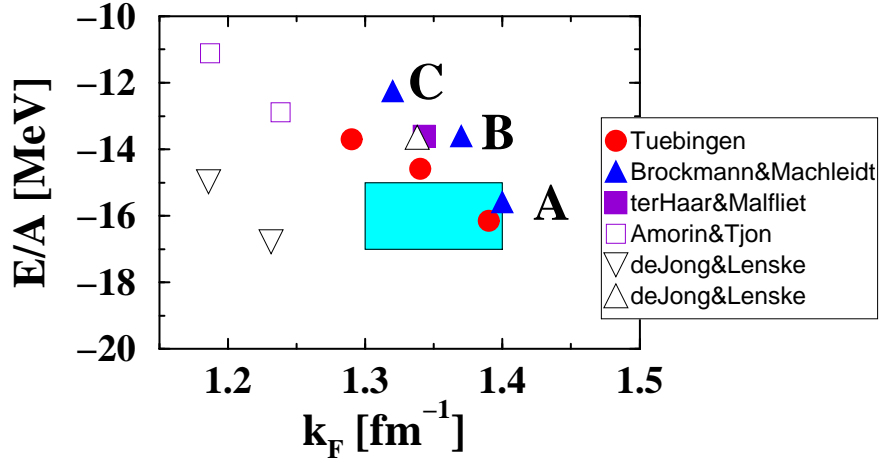}
\end{center}
\caption{Saturation points of nuclear matter, obtained 
with different OBEPs and within different approximation schemes. 
Full symbols correspond to standard relativistic Brueckner 
calculations, open symbols represent calculations which 
include Dirac sea contributions. The shaded area represents the 
empirical region of saturation.
}
\label{eos2_fig}
\end{figure}

In Fig. \ref{eos2_fig} we summarize the saturation points for 
iso-spin symmetric nuclear matter using different OBEPs 
as well as different approximation schemes. The 
saturation points for the two possible 
representations (\ref{tmatps},\ref{tmatpv}) for the subtracted T-matrix 
are very similar \cite{boelting99}. Therefore in the following 
I will  consider the $ps$ representation of the subtracted 
T-matrix as the optimal choice and compare this treatment with 
other works. With Bonn A one can reproduce the 
empirical saturation point
of nuclear matter, shown as shaded region in the figure.
The other Bonn potentials give less binding although the 
saturation density is always close to the empirically known value.
Compared to the calculations of Brockmann and Machleidt 
\cite{brockmann90} our {\it Coester-line} is slightly shifted towards the 
empirical region which indicates that a refined treatment of the 
T-matrix leads to an enhancement
of the binding energy connected with a reduced saturation density.
In addition the result of ter Haar and Malfliet \cite{terhaar87} based on 
the Groningen OBEP is shown. All results 
were obtained in the {\it no sea approximation}.

A second group of points (open symbols in Fig. \ref{eos2_fig}) includes
explicitely negative energy states in the formalism. Here one has to
keep in mind that standard OBEPs are derived within the 
no sea approximation and should therefore be used with caution in 
such calculations. The open triangle shows the result for Bonn C 
from \cite{dejong98} which is close to the corresponding Bonn C values of 
the standard treatment. Significant differences occur when OBEPs 
are used which were derived in full Dirac space. This is the case for the 
calculations of Amorin and Tjon \cite{amorin92} based on the 
Utrecht potential \cite{utrecht} and those of de Jong and Lenske 
\cite{dejong98} based on the Gross OBEP \cite{gross} (open
triangles down). These results group at low densities which 
indicates strong additional repulsive components in the NN 
interaction resulting in a rather stiff EOS. 
As discussed in \cite{dejong98} the coupling to the Dirac-sea, 
in particular the coupling to nucleon-antinucleon pairs ($Z$-graphs) 
generates a strong dynamical repulsion. 
In contrast to standard OBEPs, where repulsive 
Dirac sea contributions are effectively absorbed in a large $\omega$ 
coupling constant, OBEPs in full 
Dirac space generate such contributions dynamically.  
This is also reflected in a significantly reduced $\omega$ 
coupling constant. In nuclear matter the Dirac sea contributions
experience a medium dependence, primarily through the reduction of
the mass gap. Consequently, the $Z$-graph contributions are strongly 
enhanced at high densities which is the main source for the large 
repulsion observed in  \cite{amorin92,dejong98}. 
Unfortunately, many-body calculations in full Dirac space 
show a strong sensitivity on off-shell effects, i.e. the corresponding 
form factors and to the three-dimensional reduction scheme of the 
BS-equation \cite{dejong98}. Here 
certainly more efforts would be needed to control 
the influence of the Dirac sea in the many-body dynamics with 
higher accuracy. It should be noticed that in non-relativistic
treatments Dirac sea contributions can be accounted for on the level of
three-body forces (see discussion below).
\subsection{The Role of Correlations}
In order to examine the role of correlations it is instructive to 
compare the full DBHF theory to the mean field picture. 
In relativistic mean field theory (MFT) 
\cite{serot86,ring96} saturation occurs generally through the interplay
between the large attractive scalar field $\Sigs$, generated by the
$\sigma$-meson, and the repulsive vector field $\Sigo$ originating from the
$\omega$-meson. In MFT the vector field grows linear with density while 
the scalar field saturates at large densities which leads finally to
saturation. 
\begin{figure}
\begin{center}
\leavevmode
\includegraphics[width=0.8\textwidth]{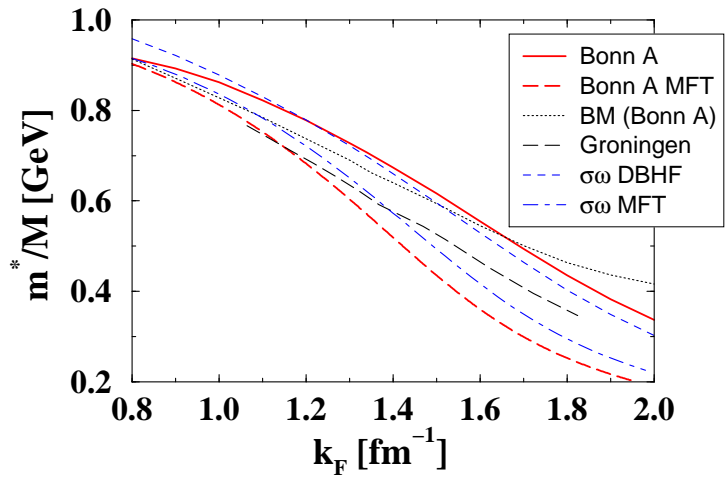}
\end{center}
\caption{The effective nucleon mass in the DBHF approach, using different 
NN interactions and approximation schemes, is compared to mean field 
calculations. 
}
\label{mstar_fig}
\end{figure}

As an illustrative example Fig. \ref{mstar_fig} shows the density 
dependence of the effective nucleon mass $\mst$ for various models: 
We compare the present DBHF result (Bonn A, $ps$ for the subtracted
T-matrix) to the mean field result for Bonn A. In the latter only 
$\sigma$ and $\omega$ mesons contribute, exchange terms and
contributions from other mesons vanish in iso-spin saturated nuclear
matter \cite{serot86,ring96}. In addition DBHF results from other 
groups are shown, i.e. those of \cite{brockmann90}
(BM), a more recent calculation of the Groningen group \cite{boersma94}, 
and the original calculation of Horowitz and Serot (HS) 
\cite{horowitz87} using only $\sigma$ and $\omega$ exchange. The
corresponding MFT result is shown as
well. In the latter two calculations the coupling constants of QHD-I
\cite{serot86}, adjusted to nuclear matter instead to NN scattering, 
have been used. Remarkable is that all 
calculations, though partially based on quite different models, lead
to the same qualitative behavior. This indicates that 
the decrease of $\mst$ and
its tendency to saturate at high densities is dictated by relativistic
dynamics. The different DBHF calculations lie thereby within a band of
about 100 MeV which is set by the usage of different OBEPs and
different approximation schemes. The reason for this common behavior
of $\mst$ is easy to understand: 

From eq. (\ref{self1}) it follows that the scalar self-energy
(\ref{trace1}) is determined by 
\beq
\Sigs ({\bf k},\kf)= \frac{4}{(2\pi)^3} \int d^3{\bf q}
\frac{\tilde{m}^*_F}{\tilde{E}^*({\bf q})}
{\theta(\kf-|{\bf q}|)} F_{\rm S}({\bf k},{\bf q};\kf)~~.
\label{sigmf}
\eeq
In MFT the scalar amplitude $F_{\rm S}$ has to be replaced by the 
corresponding coupling constant of the scalar meson
$(g_\sigma/m_\sigma)^2$. Eq. (\ref{sigmf}) represents  then 
nothing else than the self-consistency equation 
for the effective mass
\beqa
\Sigs (\kf) = -\frac{g_{\sigma}^2}{m_{\sigma}^2}\rho_{\rm S} =
-\frac{g_{\sigma}^2}{m_{\sigma}^2} \frac{4}{(2\pi)^3} \int d^3{\bf q}
\frac{\tilde{m}^*_F}{\tilde{E}^*({\bf q})}
\label{rhos}
\eeqa
which automatically leads to a saturating behavior for the 
attractive scalar field at large densities \cite{serot86}. The
momentum dependence of the T-matrix elements is generally moderate 
\cite{schiller}. This explains also why DBHF results can well be
approximated within density dependent mean field theory 
\cite{fuchs95,lenske03} which means to replace $F_{\rm S}$ in
(\ref{sigmf}) by an average value ${\overline F}_{\rm S}(\kf)$. Also 
the variation of such average amplitudes with density is 
in general moderate, however,  
with a tendency to decrease with density. Hence, one leading effect 
for saturation which takes place on the scale of the large scalar 
and vector fields of a few hundred MeV is present in full DBHF theory
as well as in MFT. 

This does, however, not mean that the saturation mechanism 
is dominated by the mean field or Hartree contribution 
and exchange terms and higher correlations play only a minor
role. 
\begin{figure}
\begin{center}
\leavevmode
\includegraphics[width=0.8\textwidth]{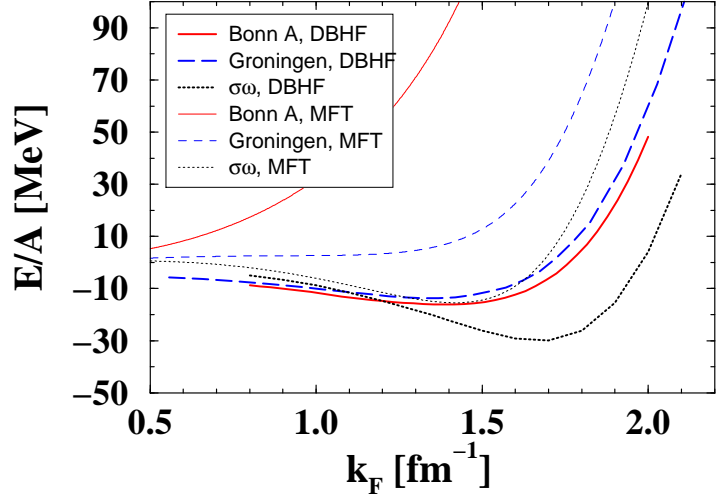}
\end{center}
\caption{The EOS obtained for various interactions in the DBHF 
approach is compared to mean field theory (MFT).
}
\label{eosmft_fig}
\end{figure}
The role of correlations can most easily be understood at the level 
of the two-nucleon wavefunction. As discussed e.g. in detail in 
\cite{muether00,schiller} and indicated in Fig. \ref{fig1},
correlations suppress the relative wavefunction at short distances. 
This reduces the short- and medium-range attraction of the
$\sigma$-meson but even more effectively the short-range repulsion of
the $\omega$-meson. Correlation lead therefore, first of all, to a general
reduction of the magnitude of the self-energies and, secondly, to
a significant reduction of the repulsive components of the 
interaction. As can be seen from Fig. \ref{eosmft_fig}, at saturation 
density the scalar field is reduced by about 100 MeV compared to MFT 
and even more at higher densities. The same holds for the vector 
field. For realistic OBEPs 
the quenching of the repulsive $\omega$ exchange is essential for the 
saturation mechanism. In a pure  mean field 
picture the system turns e.g. out to be unbound for Bonn A. When 
couplings are already adjusted to nuclear matter in MFT, as done 
in QHD-I where $g_{\omega}^2/4\pi = 10.84$ is about half of the Bonn A value 
$g_{\omega}^2/4\pi = 20$ while $g_{\sigma}^2$
is approximately the same, the higher order correlations lead to a 
significant softening of the EOS  and shift the saturation point to 
higher densities \cite{horowitz87}. This behavior is illustrated in
Fig. \ref{eosmft_fig} where DBHF results are compared to mean field 
calculations. The latter ones contain only contributions from 
$\sigma$ and $\omega$ exchange, however, with the coupling strengths of
the corresponding OBEPs Bonn A and Groningen. The calculations denoted 
by $\sigma\omega$ in Fig. \ref{eosmft_fig} are in both cases based on 
$\sigma\omega$ exchange only, however, now with the corresponding
couplings of QHD-I. 
\subsection{Role of the Pauli Operator}
Another important in-medium effect is represented by the Pauli operator 
$Q$ which projects the intermediate states in the BS-equation
(\ref{BSeq}) onto unoccupied phase space areas. The influence of Pauli 
blocking on the dynamics is most clearly seen if one considers
directly matrix elements or, respectively in-medium cross sections 
\cite{thm87b,lima93,alm94,fuchs01}. The differential on-shell cross
section ($p=q$) is given by 
\beq
d\sigma = \frac{(\mst)^4}{{\tilde s}^* 4 \pi^2} | {\hat T} (q,q,\theta) |^2 
d\Omega~.
\label{cross1}
\eeq
The squared matrix elements are obtained by the summation over 6  
helicity helicity matrix elements (5 of them are independent) in the 
partial wave basis \cite{erkelenz74,bonn,fuchs01}. 
From (\ref{cross1}) one sees first of all that, compared to free 
scattering in the medium appears a suppression factor $(\mst/M)^2$ which is
solely   due to kinematics. Furthermore, the Pauli operator modifies the optical
theorem \cite{alm94,fuchs01} and damps in particular the imaginary 
part of the T-matrix which is directly proportional to $Q$ (using 
R-matrix theory). For details see \cite{fuchs01}. 
\begin{figure}
\begin{center}
\leavevmode
\includegraphics[width=0.6\textwidth]{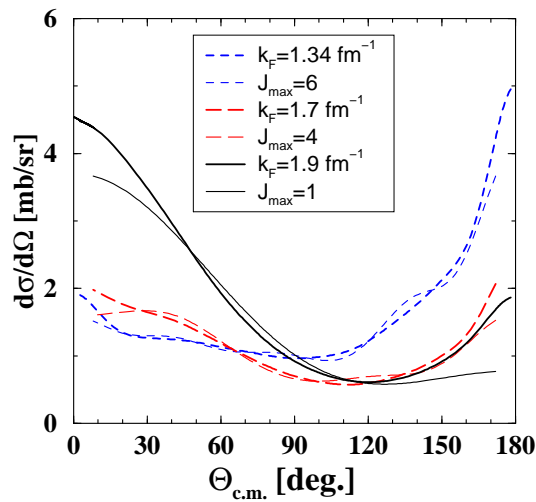}
\end{center}
\caption{Differential in-medium neutron-proton cross section for  
various densities at fixed laboratory energy of 250 MeV. The full 
results (thick lines) include partial waves up to $J=12$ while 
the thin lines were obtained truncating higher partial waves. 
}
\label{npsig_fig}
\end{figure}

Another important Pauli effect is the suppression of higher partial 
waves in the two-body correlations. This effect can be estimated from
Fig. \ref{npsig_fig} where the medium dependence of the 
differential neutron-proton cross section (at $E_{\rm lab}=250$ MeV,
using Bonn A) is 
displayed. The Fermi momenta correspond roughly to densities 1,2, and 3
(in terms of $\rho_0=0.1625~{\rm fm}^{-3}$). The free $np$ cross section is 
strongly forward-backward peaked. At moderate densities 
the presence of the medium tends 
to make the $np$ differential cross section more isotropic. 
At backward angles the cross section are decreasing with density. 
At forward angles the behavior is more complicated: At moderate 
densities the cross section is reduced but at high 
densities  ($\rho=2~{\rm and}~3~\rho_0$) a strong enhancement of the 
forward scattering amplitude can be observed. Similar results 
have been obtained by Li and Machleidt \cite{lima93}. 
At 3$\rho_0$ the cross section turns out to be again highly 
anisotropic and to be dominated by a $p$-wave component. There 
occurs generally a suppression of higher partial waves with 
increasing density: At $\rho_0$ one needs partial waves up to 
at least $J\leq 6$ to approximate the full result ($J=12$), at $2\rho_0$ 
the partial waves $J\leq 4$ are almost sufficient 
and at $3\rho_0$ the behavior is 
dominated by  $s+p$-waves. 

It is quite natural that Pauli blocking is most efficient for the 
low momentum components of the interaction, 
generated mainly by iterated $\pi$-exchange, 
while high momentum components from iterated heavy meson exchange
($\sigma,\omega$) are much less affected. 

Another effect which is closely related to the Pauli operator is a
possible onset of superfluidity at low densities. E.g. in the finite 
temperature approach of Alm et al. \cite{alm94} a critical enhancement 
of the $np$ cross section at low densities has been observed which 
was attributed to the onset of superfluidity. Crucial for such a  
superfluid state are contributions from hole-hole 
scattering in the Pauli operator ($Q= (1-f-f)$) 
which are absent in the standard Brueckner approach ($Q=(1-f)(1-f)$). 
However, as discussed in 
\cite{vonderfecht} a signature of a bound pair state can appear 
at low densities even when hole-hole scattering is neglected in the 
Pauli operator. In \cite{fuchs01} such an resonance like 
enhancement was seen in the 
amplitudes which correspond to the quantum numbers of the deuteron, 
i.e. the $^{3}S_1$, $^{3}D_1$ and the  $^{3}S_1$-$^{3}D_1$ transition 
channels. Therefore the low density enhancement of the $np$ cross section 
can be interpreted as a precursor of a superfluid state. The same
effect has been discussed in \cite{vonderfecht}.

\section{Relativistic versus Non-Relativistic BHF}
In contrast to relativistic DBHF calculations which came up in the late
80ies non-relativistic BHF theory has already almost half a century's
history. The first numerical calculations for nuclear matter were carried
out by Brueckner and Gammel in 1958 \cite{gammel}. Despite strong 
efforts invested in the development of methods to solve the 
Bethe-Goldstone (BG) equation, the non-relativistic counterpart of the 
BS equation, it turned out that, although such calculations were able to
describe the nuclear saturation mechanism qualitatively, they failed 
quantitatively. Systematic studies for a large variety of NN
interactions showed that saturation points were always allocated on a
so-called {\it Coester-line} in the $E/A-\rho$ plane which does not
meet the empirical region of saturation. In particular modern OBEPs
lead to strong over-binding and too large saturation densities where 
relativistic calculations do a much better job. 
Several reasons have been discussed in the literature in order to
explain the success of the relativistic treatment. In the following I
will recapitulate the main arguments for this difference. 

\subsection{Continuous Choice versus Gap Choice} 
Brueckner theory converges in terms
of the hole-line expansion (for a recent review see
\cite{muether00}). In lowest order Brueckner theory (2 hole-lines) the
effective 2-particle propagator in the BG-equation
leads to a gap at the Fermi surface
\beq
iG_{12} = \frac{\overline Q}{\epsilon({\bf q}) - E({\bf k}) }
\label{gap}
\eeq
with $  \epsilon({\bf q}) = U(\kf) + \frac{{\bf q}^2}{2m^*}$ the single 
particle energies below the Fermi momentum (starting energy) and 
$E({\bf k}) = \frac{{\bf k}^2}{2M}$ the energy of the intermediate
states above $\kf$. From
(\ref{thompsoneq}) it is evident that the relativistic propagator does 
not contain such a gap. The continuous choice advocated by the Liege 
group \cite{liege} assumes the single 
particle potential to be valid also above $\kf$. This is in line with the 
relativistic propagator where fields are present below and above the 
Fermi momentum. Compared to the gap (or standard) choice 
the continuous choice shifts the Coester-line
significantly towards the empirical region \cite{baldo95,banerjee02}. It was
further shown by the Catania group \cite{baldo98} that at the
3-hole-line level both choices lead to almost identical results.  
In the continuous choice already lowest order Brueckner theory 
(2-hole-lines) is very close to the result of the 3-hole-line
expansion which suggests a faster convergence of the continuous
choice. 
\begin{figure}
\begin{center}
\leavevmode
\includegraphics[width=0.8\textwidth]{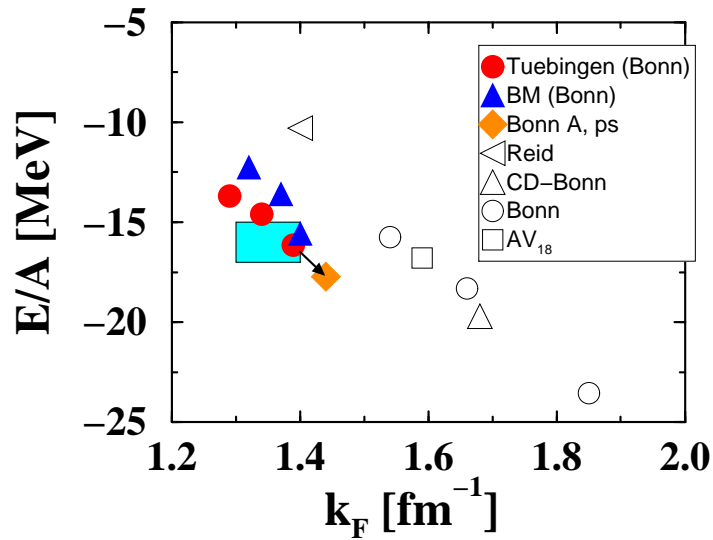}
\end{center}
\caption{Saturation points of relativistic (full symbols) 
versus non-relativistic (open symbols) BHF calculations (continuous choice). 
}
\label{eos3_fig}
\end{figure}
\subsection{Dirac Effects and Quenching of the Tensor Forces}
The saturation mechanisms in relativistic and non-relativistic
theories are quite different. In relativistic MFT the vector field grows 
linear with density while 
the scalar field saturates at large densities. In MFT this is the
essential effect which leads to
saturation. As discussed above, the density dependence of the
scalar and vector DBHF self-energy is similar to MFT, 
however, exchange contributions and 
correlations lead a significant reduction of their absolute 
magnitude. A pure mean field picture 
works when $\sigma$ and $\omega$
couplings are adjusted to nuclear matter, however, when fitted to free
scattering data no saturation may occur at the mean field level, 
depending on the choice of the interaction. Exchange contributions 
and correlations are of crucial importance 
in order to obtain the saturation point 
at a scale which is set by the binding energy. 
A genuine feature of relativity, present in MFT 
as well as in DBHF, is the 
presence of large scalar and vector fields of a few 
hundred MeV size and a strongly decreasing effective nucleon 
mass with the tendency to saturate at high densities.  

In the non-relativistic case the situation is quite different: 
The saturation mechanism takes place exclusively on the scale of the 
binding energy, i.e. a few ten MeV. It 
cannot be understood by the absence of a tensor
force \cite{bethe}. In particular the second order OPEP is large and
attractive at high densities and its interplay with Pauli-blocking
leads finally to saturation. Relativistically the interaction $V$ receives a
density dependence since matrix elements of $V$ are built between 
in-medium spinors (\ref{spinor}). As discussed in \cite{banerjee98}
the tensor interactions $V_\pi,V_\eta$ and the Pauli part of $V_\rho$ 
experience an additional reduction by the factor $(\mst/M)^2$. This 
reduction is at a first glance not completely obvious since it enters through the 
momentum dependence of  second and higher order 
pseudo-vector OPE (and similar for $\eta$ and $\rho$) where the range of the  
intermediate momentum ${\bf k}$ is controlled by $\mst$. Consequently, 
the OPEP contribution to the binding energy is significantly reduced 
in the relativistic approach. Consistent with this observation 
is the present result (Figs. 
\ref{eos1_fig} and \ref{eos3_fig}) 
obtained by a pure pseudo-scalar representation of the T-matrix 
within the projection scheme. As discussed above this leads to 
stronger weights of the self-energy contributions from 
pion exchange and in turn to stronger binding at higher density. 
It should, however, be kept in mind that in this calculation 
only the representation of the
T-matrix is taken as $ps$, the $V_\pi$ itself is still used with 
$pv$ coupling.

In summary we have two genuine features of relativistic dynamics 
which are closely connected, act, however, at different scales: the 
saturation of the scalar attraction takes place on the scale of the 
large self-energy fields and the quenching of the tensor force 
at the scale of the binding energy. As argued in 
\cite{banerjee98,banerjee02} the latter effect is probably responsible 
for the improved Coester lines compared to BHF. In the language
of effective field theory it may be tempting to relate this two 
scales with chiral fluctuations on top of large background fields
originating from QCD condensates \cite{finelli}. Remarkably, such 
an approach can finally lead to very similar self-energies as the present 
DBHF calculations  \cite{finelli}.
\subsection{Resonance Degrees of Freedom and Three-Body Forces}
Since the inclusion of explicit resonance degrees of freedom (DoFs) in
the formalism is closely related to the occurrence of 3-body forces
(3-BFs) I will discuss here both aspects in combination. 

The most important resonance is of course the $\Delta$(1232)
isobar. At low and intermediate energies it provides large part of the
intermediate range attraction and generates most of the inelasticity
above the pion threshold. Intermediate $\Delta$ states appear in
elastic $NN$ scattering only in combination with at least two-isovector-meson
exchange ($\pi\pi,~\pi\rho,\dots$) and give rise to a new class of
box diagrams. As has been shown by the Bonn group \cite{bonn2} this class
of diagrams can satisfactorily be absorbed into the effective
$\sigma$-exchange. If the $\Delta$ is maintained as an explicit DoF in
$NN$ scattering, it provides additional attraction and the
corresponding $\sigma$ strength has to be readjusted. This leads e.g. 
in the work of the Groningen group to a reduction of 
$g_{\sigma}^2/4\pi=7.4$ (w/o $\Delta$) to $g_{\sigma}^2/4\pi=6.4$ 
including $\Delta$ DoFs.

\begin{figure}
\begin{center}
\leavevmode
\includegraphics[width=0.8\textwidth]{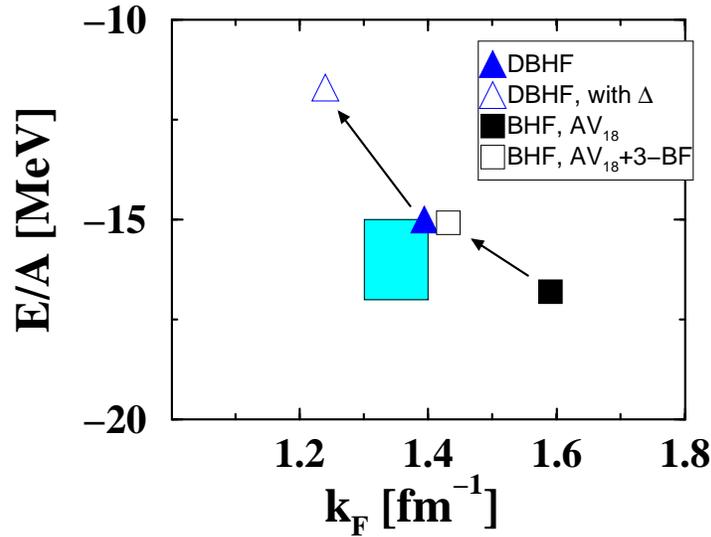}
\end{center}
\caption{Saturation points of DBHF calculations from the Groningen 
group without and including explicit $\Delta$ 
degrees of freedom \protect\cite{dejongthesis}, and of BHF calculations 
\protect\cite{zuo02} based on the $AV_{18}$ potential without and 
including 3-body forces.
}
\label{eos3bf_fig}
\end{figure}

In many-body calculations explicit $\Delta$ DoFs give rise to 
additional saturation, shifting the saturation point away 
from the empirical region. This  can be understood in the 
following way \cite{machleidt01}: 
while the elementary $\sigma NN$ vertex is not modified in the medium,  
corresponding box diagrams with intermediate $\Delta$ and nucleon 
lines are affected. Dressing of the
propagators and Pauli blocking of the nucleon state suppresses their 
contribution. Iterated to all orders, the maintenance of explicit 
$\Delta$ DoFs (instead of a stronger $\sigma$ exchange) results therefore 
in less attraction. Quantitatively this effect has in detail 
been studied by ter Haar and Malfliet \cite{terhaar87}. 
Fig. \ref{eos3bf_fig} shows the result of a more refined calculation from
the Groningen group \cite{dejongthesis} which includes also the
$\Delta$ self-energy.  As one sees, the loss of binding
energy is quite substantial. 
As pointed out in Refs. \cite{muether00,machleidt01} the
inclusion of non-nucleonic degrees of freedom has to be
performed with caution: Freezing out resonance DoFs 
generates automatically a class of three-body forces which contains 
nucleon-resonance excitations. 3-BFs with intermediate $\Delta$ 
excitations provide again a strong intermediate range attraction,  
$N^*(1440)$ excitations lead to small net repulsion \cite{coon95}. 
Hence there exist strong
cancellation effects between the repulsion due to box diagrams 
and contributions from 3-BFs. Such ring
type diagrams of third and forth order in the hole-line expansion
have e.g. been conducted in \cite{dickhoff82}. A
consistent treatment requires therefore to consider non-nucleonic DoFs
and many-body forces on the same footing \cite{machleidt01}. Many-body 
forces which are exclusively based on nucleon degrees of freedom can
systematically be generated within chiral perturbation theory. Next to
leading order all 3-BFs cancel \cite{weinberg90} 
while non-vanishing contributions appear at NNLO \cite{klock94}.  

Another class of 3-BFs which has extensively been studied within
non-relativistic BHF involves virtual excitations of
nucleon-antinucleon pairs. Such Z-graphs  are
in net repulsive \cite{coon95} and lead to a shift of the saturation point away
from the  non-relativistic Coester line towards its relativistic
counterpart where the DBHF results are allocated. The calculation
shown in Fig. \ref{eos3bf_fig} includes both, ${\bar N}N$ and 
as well as nucleon-resonance
excitations \cite{zuo02}. 

It is often argued that in non-relativistic
treatments 3-BFs play in some sense an equivalent role as the
dressing of the two-body interaction by in-medium spinors in 
Dirac phenomenology. Both mechanisms lead indeed to an effective density
dependent two-body interaction $V$ which is, however, of different
origin. In the medium 3-BFs can be considered as a renormalization of 
the meson vertices and propagators. Z-graphs are explicitely
included when DBHF calculations are performed in full Dirac space,  
in the no sea approximation they are in
some way effectively included through the usage of OBEPs with large
$\omega$ couplings. 

\section{Summary}
An overview on the present status of relativistic
Brueckner calculations for the nuclear many-body problem was given. 
Using modern one-boson-exchange potentials such calculations provide a 
qualitatively satisfying - and parameter free - description of the 
nuclear saturation mechanism. Concerning the extractions of the
precise nuclear self-energy and its Lorentz structure there arise
on-shell ambiguities due to lack of information on two-body matrix
elements in full Dirac space when the approach is restricted to the 
positive energy sector (no sea approximation). A method to minimize 
the corresponding uncertainties was discussed. 

Similar to relativistic mean field theory, Dirac phenomenology
together with the structure of the NN interaction extracted from free 
scattering implies the existence of large scalar and vector
fields. However, exchange contributions and higher order correlations
reduce the magnitude of these fields compared to MFT and are essential
for a quantitative saturation mechanism. When calculations are
performed in full Dirac space part of the repulsion is generated from 
sea excitations which requires to renormalize the NN
potentials. However, in the latter case the many-body dynamics can 
presently not be controlled with the same accuracy as in the standard 
approach based on the no sea approximation. In non-relativistic BHF 
nucleon-antinucleon excitations can be accounted on the level of
three-body forces which leads to qualitatively similar results as 
in relativistic approach with solely two-body interactions. However, 
a consistent treatment of 3-BFs is a subtle problem which is 
closely connected to the introduction  of non-nucleonic degrees of
freedom, i.e. nuclear resonances. A future perspective would be the 
application of chiral NN potentials \cite{entem03} where 3-BFs can 
consistently be by power counting. Another somewhat 
complementary challenge, relevant for the application to relativistic heavy ion
reactions, is an extension to higher energies, using thereby high 
precision potentials above the pion threshold \cite{geramb01} .
\\
\noindent 
{\bf Acknowledgments}\\
\noindent 
The author acknowledges valuable discussions 
with H. Lenske and W. Weise.


\printindex
\end{document}